# The significance of the Schott energy for energy-momentum conservation of a radiating charge obeying the Lorentz-Abraham-Dirac equation

Ø. Grøn

Oslo University College, Faculty of Engineering, P. O. Box 4, St. Olavs Pl., N-0130 Oslo, Norway

**Abstract**

It is demonstrated that energy and momentum is conserved during runaway motion of a radiating charge, and during free fall of a charge in a field of gravity. This does not mean that runaway motion is really happening. It may be an unphysical solution of the Lorentz-Abraham-Dirac (LAD) equation of motion of a radiating charge. However it demonstrates the consistency of classical electrodynamics, including the LAD equation which is deduced from Maxwell's equations and the principle of energy-momentum conservation applied to a radiating charge and its electromagnetic field. The decisive role of the Schott energy in this connection is made clear, and an answer is given to the question: What sort of energy is the Schott energy and where is it found? It is the part of the electromagnetic field energy which is proportional to (minus) the scalar product of the velocity and acceleration of a moving accelerated charged particle. In the case of the electromagnetic field of a point charge it is localized at the particle. This energy is negative if the acceleration is in the same direction as the velocity and positive if it is in the opposite direction. During runaway motion the Schott energy becomes more and more negative and in the case of a charged particle with finite extension, it is localized in a region with increasing extension surrounding the particle. The Schott energy provides both the radiated energy and the increase of kinetic energy. Also it is pointed out that a proton and a neutron fall with the same acceleration in a uniform gravitational field although the proton radiates and the neutron does not. Again the radiation energy comes from the Schott energy.

### 1. Introduction

D. J. Griffith et al.[1] have recently presented an interesting comparison of the Abraham-Lorentz-Dirac (LAD) versus the Landau-Lifshitz equation of motion of a radiating charge. However, in their introduction they write two sentences that may be somewhat misleading because they are not generally valid. In connection with the problem that the LAD-equation permits a charge which is not acted upon by external forces, to have an accelerated runaway motion, they write: "Runaways violate conservation of energy". Furthermore, having pointed out that the power $P_L$ radiated by a particle with charge $q$ and acceleration $a$ is given by the Larmor formula,

$$P_L = k_L a^2 \ , \qquad (1)$$

where $k_L = q^2/6\pi\varepsilon_0 c^3$ in S.I.-units and $k_L = (2/3c^3)q^2$ in cgs-units, they write that "the energy lost to radiation comes at the expense of the particle's kinetic energy – a charged particle accelerates less than its uncharged twin."

We shall here analyze two examples, one showing that runaway motion conserves energy and momentum when the Schott energy is taken into account, and the other showing that a charge falling freely in a gravitational field radiates according to the Larmor formula, and yet it moves in the same way as a freely falling neutral particle. The radiation energy does not come at the expense of the particle's kinetic energy, but at expense of a decrease of its Schott energy.

First we shall present the role played by the Schott energy in the dynamics of charged particles, starting with a simple, non-relativistic description, and proceeding with the covariant, relativistic equation of motion. Then we shall explain what sort of energy the Schott energy is and where it is found. Runaway motion is described in section 4, and it is shown that energy-momentum is conserved during this motion, and how this can be understood by taking account of the Schott energy. Finally we consider a radiating charged particle falling freely in the Rindler space of a uniformly accelerated reference frame. It is demonstrated that the charged particle falls with the same acceleration as a neutral particle although the charged particle radiates and the neutral one does not. Also it is shown that the radiation energy comes from a corresponding decrease of the Schott energy of the charge's electromagnetic field.

### 2. The dynamics of a charged particle

The analysis of the energy-momentum balance of a radiating charge is usually based on the equation of motion of a point charge. The non-relativistic version of the equation was discussed already more than a hundred years ago by H. A. Lorentz.[2] The relativistic generalization of the equation was originally found by M. Abraham[3] in 1905 and re-derived in 1909 by M. Von Laue[4] who Lorentz transformed the non-relativistic equation from the instantaneous rest frame of the charge to an arbitrary inertial frame. A new deduction of the Lorentz covariant equation of motion was given by P. A. M. Dirac in 1938.[5] This equation is therefore called the Lorentz-Abraham-Dirac equation, or for short the LAD-equation. A particularly interesting feature about Dirac's deduction is that it establishes a connection between Maxwell's equations and the equation of motion for a charged particle. It shows that the presence of the Abraham four-vector in the equation of motion comes from conservation of energy and momentum for a closed system consisting of a charge and its electromagnetic field.

### 2.1. The non-relativistic equation of motion of a radiating charge

In the non-relativistic limit the equation of motion of a radiating charge, $q$, with mass $m_0$, acted upon by an external force, $\mathbf{f}_{ext}$, takes the form

$$m_0 \ddot{\mathbf{r}} = \mathbf{f}_{ext} + m_0 \tau_0 \dddot{\mathbf{r}} \quad , \quad \tau_0 = \frac{q^2}{6\pi\epsilon_0 m_0 c^3} = \frac{k_L}{m_0} \quad , \tag{2}$$

where the dot denotes differentiation with respect to the (Newtonian) time. Here $\tau_0$ is or the same order of magnitude as the time taken by light to move a distance equal to the classical electron radius, i.e. $\tau_0 \approx 10^{-23}$ seconds. The general solution of the equation is

$$\ddot{\mathbf{R}}(T) = e^{T/\tau}\left[\ddot{\mathbf{R}}(0) - \frac{1}{m\tau_0}\int_0^T e^{-T'/\tau_0}\mathbf{f}_{ext}(T')dT'\right] \quad . \tag{3}$$

Hence the charge performs a runaway motion unless on chooses the initial condition

$$m\tau_0\ddot{\mathbf{R}}(0) = \int_0^\infty e^{-T'/\tau_0}\mathbf{f}_{ext}(T')dT' \quad . \tag{4}$$

However, by combining Eq.(3) and Eq.(4) one obtains[6]

$$m\ddot{\mathbf{R}}(T) = \int_0^\infty e^{-s}\mathbf{f}_{ext}(T + \tau_0 s)ds \quad . \tag{5}$$

This equation shows that the acceleration of the charge at a point of time $T$ is determined by the future force, weighted by a decreasing exponential factor with value 1 at the time $T$, and a time constant $\tau_0$, i.e. there is pre-acceleration.

In his discussion of Eq.(2) Lorentz[1] writes: "In many cases the new force represented by the second term in Eq.(2) may be termed a *resistance* to the motion. This is seen if we calculate the work of the force during an interval of time extending from $T = T_1$ to $T = T_2$. The result is

$$\frac{2q^2}{3}\int_{T_1}^{T_2} \dot{\mathbf{a}} \cdot \mathbf{v}\, dT = \frac{2q^2}{3}[\dot{\mathbf{a}} \cdot \mathbf{v}]_{T_1}^{T_2} - \frac{2q^2}{3}\int_{T_1}^{T_2} \mathbf{a} \cdot \mathbf{a}\, dT \quad . \tag{6}$$

Here the first term disappears if, in the case of periodic motion, the integration is extended to a full period, and also if at the instants $T_1$ and $T_2$ either the velocity or the derivative of the acceleration is zero. Whenever the above formula reduces to the last term, the work of the force is seen to be negative, so that the name of resistance is then justly applied.

P. Yi[7] gives the following interpretation: "The total energy of the system may be split into three pieces: the kinetic energy of the charged particle, the radiation energy, and the electromagnetic energy of the Coulomb field. In effect, the last acts as a sort of energy reservoir that mediates the energy transfer from the first to the second and *in the special case of uniform acceleration provides all the radiation energy without extracting any from the charged particle.*"

**2.2. The relativistic equation of motion of a radiating charge**

The relativistic equation of motion of a particle with rest mass $m_0$ and charge $q$ (the LAD-equation) may be written[8]

$$F^\mu_{ext} + \Gamma^\mu = m_0 \dot{U}^\mu \quad , \tag{7}$$

where

$$\Gamma^\mu \equiv k_L\big(\dot{A}^\mu - A^\alpha A_\alpha U^\mu\big) \quad , \tag{8}$$

and the dot denotes differentiation with respect to the proper time of the particle. Here $F^\mu_{ext}$ is the external force acting upon the particle, $U^\mu$ is its 4-velocity and $A^\mu$ its 4-acceleration. (Capital letters shall be used for 4-vector components referring to an inertial frame, and units are used so that $c = 1$.)

The vector with components $\Gamma^\mu$ is called the *Abraham 4-force* and is given by

$$\Gamma^\mu = k_L \gamma(\mathbf{v} \cdot \mathbf{\Gamma},\ \mathbf{\Gamma}) \quad , \tag{9}$$

where $k_L \mathbf{\Gamma} = m_0 \tau_0 \mathbf{\Gamma}$ is the three-dimensional force called the *field reaction force*[9], $\mathbf{v}$ is the ordinary velocity of the particle, and $\gamma = (1 - v^2)^{-1/2}$. In an inertial reference frame the Abraham 4-force may be written

$$\Gamma^\mu = k_L \gamma(\mathbf{v} \cdot \dot{\mathbf{g}},\ \dot{\mathbf{g}}) \quad . \tag{10}$$

where $g = (A_\alpha A^\alpha)^{1/2}$ is the proper acceleration of the charged particle in the inertial frame. Hence

$$\mathbf{\Gamma} = \dot{\mathbf{g}} \ . \tag{11}$$

In flat spacetime there exist global inertial frames. However, in curved spacetime there are only *local* inertial frames. They are freely falling. Then **g** is the acceleration of the particle in a freely falling frame in which the particle is instantaneously at rest. In such a frame a freely falling particle has no acceleration. Hence, a particle falling freely in a gravitational field has vanishing 4-acceleration. From Eq.(8) and Eq.(11) is seen that for such a particle the Abraham 4-force vanishes. This is also valid for a charged particle emitting radiation while it falls. This case shall be treated in more detail in section 5.

According to the Lorentz covariant Larmor formula, valid with reference to inertial systems, the energy radiated by the particle per unit time is, (using the sign convention that the signature of the metric is +2),

$$P_L = k_L A^\alpha A_\alpha = k_L g^2 \ . \tag{12}$$

The radiated momentum per unit proper time is

$$P_R^\mu = P_L U^\mu \ . \tag{13}$$

From the equation of motion (7) we get the energy equation

$$\gamma \mathbf{v} \cdot \mathbf{F}_{ext} = m_0 \dot{U}^0 - \Gamma^0 = m_0 \gamma^4 \mathbf{v} \cdot \mathbf{a} - \gamma \mathbf{v} \cdot \mathbf{\Gamma} = \frac{dE_K}{dT} - \gamma \mathbf{v} \cdot \mathbf{\Gamma} \ , \tag{14}$$

where $E_K = (\gamma - 1) m_0 c^2$ is the kinetic energy of the particle, and $T$ is the coordinate time in the inertial frame. Note that the energy supplied by the external force is equal to the change of the kinetic energy of the charge when the Abraham 4-force vanishes. Hence, it is tempting to conclude from the Abraham Lorentz theory, i.e. from Eq. (11) and Eq.(14), that a charge having constant acceleration does not radiate. This is, however, not the case. The power due to the field reaction force is

$$\mathbf{v} \cdot \mathbf{\Gamma} = \frac{d}{dT}(k_L \gamma^4 \mathbf{v} \cdot \mathbf{a}) - P_L = -\frac{dE_S}{dT} - \frac{dE_R}{dT} \ , \tag{15}$$

where $E_R$ is the energy of the radiation field, and $E_S$ is the Schott energy defined by,

$$E_S \equiv -k_L \gamma^4 \mathbf{v} \cdot \mathbf{a} = -k_L A^0 \ . \tag{16}$$

(This energy was called "acceleration energy" by Schott[10], but is now usually called "Schott energy".) Hence, in the case of constant acceleration, when the Abraham 4-force vanishes, the charge radiates in accordance with Larmor's formula, Eq.(12), and the rate of radiated energy is equal to minus the rate of change of the Schott energy. The energy equation may now be written

$$\frac{dW_{ext}}{dT} = \mathbf{v} \cdot \mathbf{F}_{ext} = \frac{d}{dT}(E_K + E_S + E_R) \ , \tag{17}$$

where $W_{ext}$ is the work on the particle due to the external force.

Let $\mathbf{P}_{ext}$ be the momentum delivered to the particle from the external force. Then $d\mathbf{P}_{ext}/dT = \mathbf{F}_{ext}$, and by means of Eq.(2), Eq.(3) and Eq.(8) we get,

$$\frac{d\mathbf{P}_{ext}}{dT} = \mathbf{F}_{ext} = m_0 \frac{d\mathbf{U}}{dT} - k_L\left(\frac{d\mathbf{A}}{dT} - g^2\mathbf{v}\right) = \frac{d\mathbf{P}_M}{dT} + \frac{d\mathbf{P}_S}{dT} + \frac{d\mathbf{P}_R}{dT} \ . \tag{18}$$

Thus, according to Eq.(13) and Eq.(18) the momentum of the particle takes the form

$$\mathbf{P}^\mu = \mathbf{P}_M^\mu + \mathbf{P}_S^\mu \ , \tag{19}$$

where

$$\mathbf{P}_M^\mu = m_0 U^\mu \ , \quad \mathbf{P}_S^\mu = -k_L A^\mu \tag{20}$$

are the mechanical momentum of the particle and the Schott-momentum, respectively. In addition we have the momentum of the radiation field, which is not a state function of the particle.

3. **The physical meaning of the Schott energy**

Already in 1915 Schott[10] argued that in the case of uniform acceleration

> "the energy radiated by the electron is derived entirely from its acceleration energy; there is as it were internal compensation amongst the different parts of its radiation pressure, which causes its resultant effect to vanish".

But what is the "acceleration energy", now called "the Schott energy"? Schott[10] and later Rohrlich[9] noted that there is an important difference between the radiation rate and the rate of change of the Schott energy.

"The radiation rate is always positive (or zero) and describes an *irreversible loss* of energy; the Schott energy changes in a *reversible* fashion, returning to the same value whenever the state of motion repeats itself."

Rohrlich also wrote:[11]

"If the Schott energy is expressed by the electromagnetic field, it would describe an energy content of the near field of the charged particle which can be changed *reversibly*. In periodic motion energy is borrowed, returned, and stored in the near-field during each period. Since the time of energy measurement is usually large compared to such a period only the average energy is of interest and that average of the Schott energy rate vanishes. Uniformly accelerated motion permits one to borrow energy from the near-field for large *macroscopic* time-intervals, and no averaging can be done because at no two points during the motion is the acceleration four-vector the same. Nobody has so far shown in detail just how the Schott energy occurs in the near-field, how it is stored, borrowed etc."

A step toward answering this challenge was taken by C. Teitelboim.[12] He made a Lorentz invariant separation of the field tensor of the electromagnetic field of a point charge, into two parts, $F^{\mu\nu} = F_I^{\mu\nu} + F_{II}^{\mu\nu}$ where $F_I^{\mu\nu}$ is the velocity field and $F_{II}^{\mu\nu}$ the acceleration field. Inserting these parts into the expression for the energy-momentum tensor of the electromagnetic field, Teitelboim found that the energy-momentum tensor contains terms of three types: a part $T_{I,I}^{\mu\nu}$ independent of the acceleration, a part $T_{I,II}^{\mu\nu}$ depending linearly upon the acceleration, and a part $T_{II,II}^{\mu\nu}$ depending linearly upon the square of the acceleration of the charged particle producing the fields. Teitelboim then defined $T_I^{\mu\nu} = T_{I,I}^{\mu\nu} + T_{I,II}^{\mu\nu}$ and $F_{II}^{\mu\nu} = F_{II,II}^{\mu\nu}$. The contribution of the interference between the fields I and II has been included in $T_I^{\mu\nu}$, whereas the tensor $F_{II}^{\mu\nu}$ is related only to the part of the field depending upon the square of the acceleration. Teitelboim showed that the energy-momentum associated with the field $F_{II}^{\mu\nu}$ is travelling with the speed of light. The field fronts are spheres with centres at the emission points. The four-momentum associated with $F_I^{\mu\nu}$ remains bound to the charge. Furthermore he calculated the four-momenta and their time derivatives associated with $F_I^{\mu\nu}$ and $F_{II}^{\mu\nu}$.

The results of Rohrlich and Teitelboim have been summarized by P. Pearle[13] in the following way:

"The term $\Gamma^\mu$ in the Lorentz-Dirac equation, as given in Eq.(3), is called the Abraham force. Its first term, $k_L \dot{A}^\mu$ is called the Schott term, and its second, $-k_L A^\alpha A_\alpha U^\mu$, the radiation reaction term. The zeroth component of the radiation reaction term is to be interpreted as the radiation rate. Indeed, the scalar product of this term with $U_\mu$ is the relativistic version of the Larmor formula. The spatial component of this term, proportional to $-\mathbf{v}$ like a viscous drag force, may similarly be interpreted as the radiation reaction force of the electron.

The physical meaning of the Schott term has been puzzled over for a long time. Its zero component represents a power which adds "Schott acceleration energy" to the electron and its associated electromagnetic field. The work done by an external force not only goes into electromagnetic radiation and into increasing the electron's kinetic energy, but it causes an increase in the "Schott acceleration energy" as well. This change can be ascribed to a change in the "bound" electromagnetic energy in the electron's induction field, just as the last term Eq.(15) can be ascribed to a change in the "free" electromagnetic energy in the electron's radiation field.

What meaning should be given to the Schott term? Teitelboim[12] has argued convincingly that when an electron accelerates, its near-field is modified so that a correct integration of the electromagnetic four-momentum of the electron includes not only the Coulomb 4-momentum $(q^2/8\pi\epsilon_0 r)U^\mu$, but an extra 4-momentum $-k_L A^\mu$ of the bound electromagnetic field."

It remained to obtain a more precise localization of the Schott field energy.

Rowe[14] has modified Teitelboim's separation of the energy-momentum tensor of the electromagnetic field of a point charge which he described by a delta-function, and introduced a separation into three symmetrical, divergence free parts. In order to obtain a finite expression for the localization of the Schott energy as part of the energy of the electromagnetic field of a charged particle, Eriksen and Grøn[8] applied Rowe's separation to a charged particle with a finite radius and obtained the following result. The Schott energy is inside a spherical light front S touching the front end of a moving Lorentz contracted charged particle. From Figure 1 one finds that at the point of time $T$ the radius of the light front S that represents the boundary of the distribution of the Schott energy, is

$$c(T - T_{Q2}) = \varepsilon \sqrt{\frac{1+v}{1-v}} \quad , \tag{21}$$

where the field at the light front S is produced at the retarded point of time $T_{Q2}$, $\varepsilon$, is the proper radius of the particle, and $v$ is the absolute value of its velocity. Hence, unless the velocity of the charge is close to that of light, the Schott energy is localized just outside the surface of the charged particle. The radius of the light front S increases towards infinity for the field of a charge approaching the velocity of light, for example during runaway motion which we shall now consider.

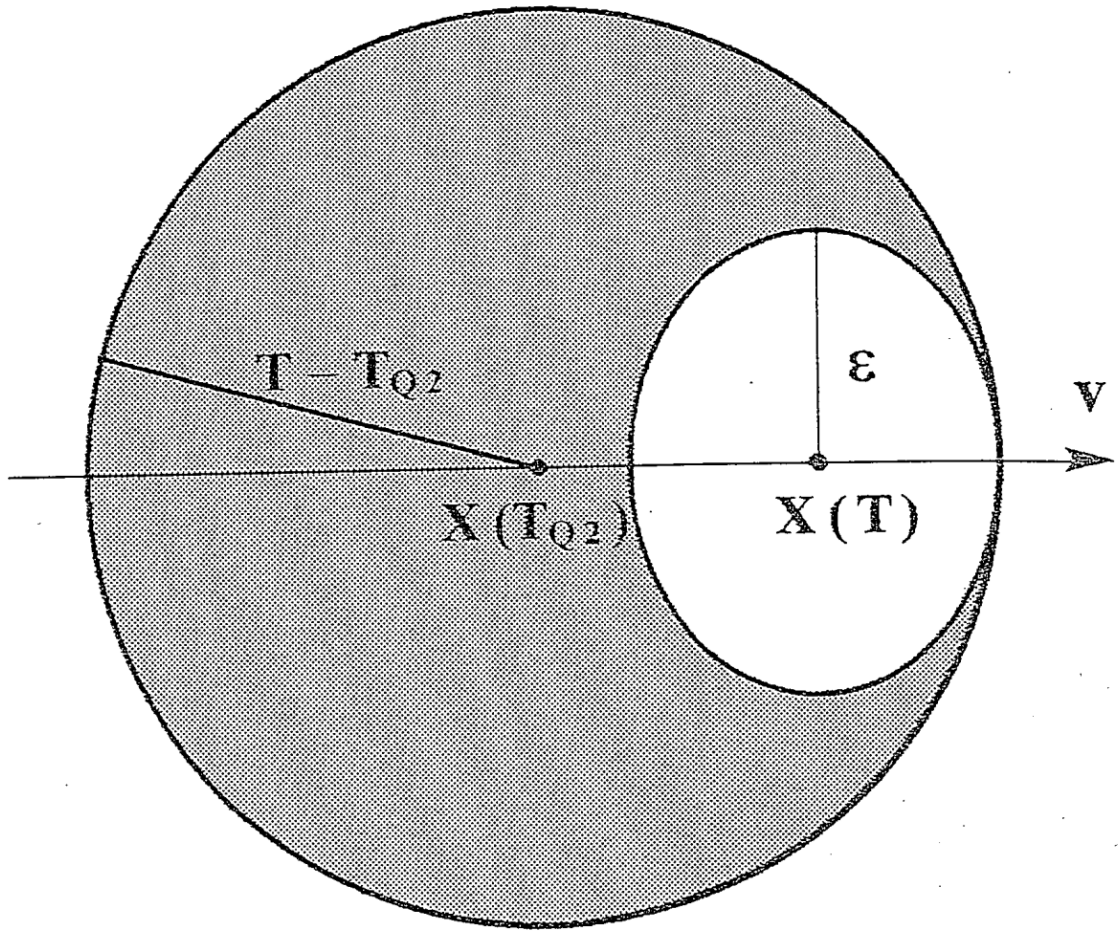

**Figure 1.** The figure shows a Lorentz contracted charged particle with proper radius $\epsilon$ moving to the right with velocity $v$. The field is observed at a point of time $T$, and at this moment the centre of the particle is at the position $X(T)$. The circle is a field front produced at the retarded point of time $T_{Q2}$ when the centre of the particle was at the position $X(T_{Q2})$. The field front is chosen such that it just touches the front of the particle. The Schott energy is localized in the shaded region between the field front and the ellipsoid representing the surface of the particle. On the figure the velocity is chosen to be $v = 0.6$.

In the deduction of the localization of the Schott energy we have not applied the equation of motion of the charge. We have only considered the field produced by the charge. Hence the deduction permits us to consider a charge with a finite proper radius. However, the LAD-equation is deduced for a point charge. So relating the Schott energy to the conservation of energy for a radiating point charge and the field it produces, we should take the limit $\epsilon \to 0$. In this limit it seems that the Schott energy is localized at the point charge. But it must be admitted that this limit seems rather unphysical, and our conclusion should rather be that this limit signals a breakdown of classical electrodynamics, or at least some sort of unsolved problem.

4. Energy conservation during runaway motion

Runaway acceleration seems to be in conflict with the conservation laws of energy and momentum. The momentum and the kinetic energy of the particle increase even when no force acts upon it. The charged particle even puts out energy in the form of radiation. Where do the energy and the momentum come from?

We shall here show that the source of energy and momentum in run-away motion is the so-called Schott energy and momentum.[15] During motion of a charge in which the velocity increases, the Schott energy has an increasingly negative value and there is an increasing Schott momentum directed oppositely to the direction of the motion of the charge.

We shall consider a charged particle performing runaway motion along the $X$-axis. Introducing the rapidity $\alpha$ of the particle its velocity and acceleration may be expressed as

$$v = \frac{dX}{dT} = \tanh\alpha \ , \quad \gamma = \cosh\alpha \ , \quad \gamma v = \sinh\alpha \ , \tag{22}$$

$$a = \frac{dv}{dT} = \frac{1}{\gamma}\frac{dv}{d\tau} = \frac{\dot\alpha}{\cosh^3\alpha} \ , \tag{23}$$

where $\tau$ is the proper time of the charged particle, and the dot denotes differentiation with respect to the proper time of the particle. The LAD equation then takes the form[16]

$$\dot\alpha - \tau_0 \ddot\alpha = F/m_0 \ , \tag{24}$$

where $F$ is the external force, and $\tau_0 = k_L/m_0$ is the time taken by a light signal to travel a distance equal to two thirds of the charged particle's classical radius. For an electron $\tau_0 = 6.2 \cdot 10^{-24} s$. Eq.(24) may be written

$$\frac{d}{d\tau}\left(e^{-\tau/\tau_0}\dot\alpha\right) = -\frac{F}{m_0 \tau_0}e^{-\tau/\tau_0} \ . \tag{25}$$

For $F = 0$, i.e. for a free particle, the solutions of the LAD-equation are

$$1) \ \dot\alpha = 0 \ , \quad i.e. \ \alpha = const. \ , \quad v = const. \tag{26}$$

which is consistent with Newton's 1. law.

$$2) \quad \dot{\alpha} = k e^{\tau/\tau_0} \quad , \quad k \neq 0 \quad , \quad i.e. \quad a \neq 0 \quad . \tag{27}$$

This is the runaway solution.

As pointed out by Dirac[5] a particle in state 1) or 2) will remain in that state as long as no external force is acting. We shall here consider a particle which is at rest, i.e. in state 1), until it is acted upon by a force $F(\tau)$ pointing in the positive $x$-direction, i.e. we consider a solution of the LAD-equation without pre-acceleration. The force is acting from $\tau_1$ to $\tau_2$. For $\tau > \tau_2$ the particle is again free.

According to the LAD-equation (24) $\dot{\alpha}$ is in the present case given by

$$\dot{\alpha}(\tau) = -\frac{e^{\tau/\tau_0}}{m_0 \tau_0} \int_{-\infty}^{\tau} F(\tau') e^{-\tau'/\tau_0} d\tau' \quad . \tag{28}$$

The integral vanishes for $\tau < \tau_1$, which gives $\dot{\alpha} = 0$ (and $\alpha = 0$). For $\tau > \tau_2$ the integral is independent of $\tau$ and we get the runaway motion Eq.(27). If the integral limit $-\infty$ in eq.(4.4) is replaced by $\infty$, pre-acceleration is introduced, and run away motion disappears.

In the following we examine Eq.(28) when the force $F$ has constant value $F_0$ between $\tau_1$ and $\tau_2$, and is equal to zero outside this interval. The solution of the equation of motion is then

$$\tau < \tau_1 \; , \qquad \dot{\alpha} = 0 \; , \; \alpha = 0 \; , \tag{29}$$

$$\tau_1 < \tau < \tau_2 \; , \qquad \dot{\alpha} = \frac{F_0}{m_0} - \frac{F_0}{m_0} e^{\frac{\tau-\tau_1}{\tau_0}} \; , \quad \alpha = \frac{F_0}{m_0}(\tau - \tau_1) - \frac{F_0 \tau_0}{m_0}\left(e^{\frac{\tau-\tau_1}{\tau_0}} - 1\right) , \tag{30}$$

$$\tau_2 < \tau \; , \; \dot{\alpha} = -\frac{F_0}{m_0}\left(e^{-\frac{\tau_1}{\tau_0}} - e^{-\frac{\tau_2}{\tau_0}}\right) e^{\frac{\tau}{\tau_0}} \; , \; \alpha = \frac{F_0}{m_0}(\tau_2 - \tau_1) - \frac{F_0 \tau_0}{m_0}\left(e^{-\frac{\tau_1}{\tau_0}} - e^{-\frac{\tau_2}{\tau_0}}\right) e^{\frac{\tau}{\tau_0}} \; . \tag{31}$$

Eq.(30) shows a strange aspect of the motion. The quantity $\dot{\alpha}$ contains two terms. The first expresses the relativistic version of Newton's 2. law, i.e. $F_0 = d(\gamma m_0 v)/dT$. However, the second term represents a runaway motion *oppositely directed* relative to the external force $F_0$, a highly unexpected mathematical result. According to Eq.(30) $\dot{\alpha}$ and $\alpha$ are oppositely directed relative to $F_0$ during the whole of the time interval $\tau_1 < \tau < \tau_2$.

At the point of time $\tau = \tau_2$,

$$\dot{\alpha}(\tau_2) = \frac{F_0}{m_0}\left(1 - e^{\frac{\tau_2-\tau_1}{\tau_0}}\right) , \tag{32}$$

$$\alpha(\tau_2) = \frac{F_0}{m_0}\left(\tau_2 - \tau_1 + \tau_0 - \tau_0 e^{\frac{\tau_2-\tau_1}{\tau_0}}\right) . \tag{33}$$

In order to simplify the expressions we let $\tau_2 \to \tau_1$ and $F_0 \to \infty$ keeping the product $(\tau_2 - \tau_1) \cdot F_0 \equiv P$ constant. We then find the limits

$$\dot{\alpha}(\tau_2) = -\frac{P}{m_0\tau_0} , \quad i.e. \quad a = -\frac{P}{m_0\tau_0} , \tag{34}$$

$$\alpha(\tau_2) = 0 , \quad i.e. \quad v = 0 . \tag{35}$$

In this limit the external force is expressed by a $\delta$-function

$$F(\tau) = \delta(\tau - \tau_1)P . \tag{36}$$

Putting $\tau_1 = 0$ we have the situation: For $\tau < 0$ the particle stays at rest. At $\tau = 0$ it is acted upon by the force

$$F = \delta(\tau)P , \tag{37}$$

giving the particle an acceleration oppositely directed relatively to the force and a vanishing initial velocity,

$$a(0) = a_0 = -\frac{P}{m_0\tau_0} , \quad v(0) = 0 . \tag{38}$$

According to Eqs.(26), (27) and (38) the motion is as follows,

$$\tau < 0 , \quad \dot{\alpha} = 0 , \quad \alpha = 0 , \tag{39}$$

$$\tau > 0 , \quad \dot{\alpha} = a_0 e^{\tau/\tau_0} , \quad \alpha = \tau_0 a_0 e^{\tau/\tau_0} - \tau_0 a_0 . \tag{40}$$

The runaway motion for $\tau > 0$ is accelerated, and the velocity $v = \tanh\alpha$ approaches the velocity of light as an unobtainable limit.

The problem is to explain how this is possible for a particle not acted upon by any external force. It must be possible to demonstrate that the energy and momentum of the particle and its electromagnetic field is conserved, and find the force causing the acceleration. Of essential importance in this connection is the Schott energy and the Schott momentum.

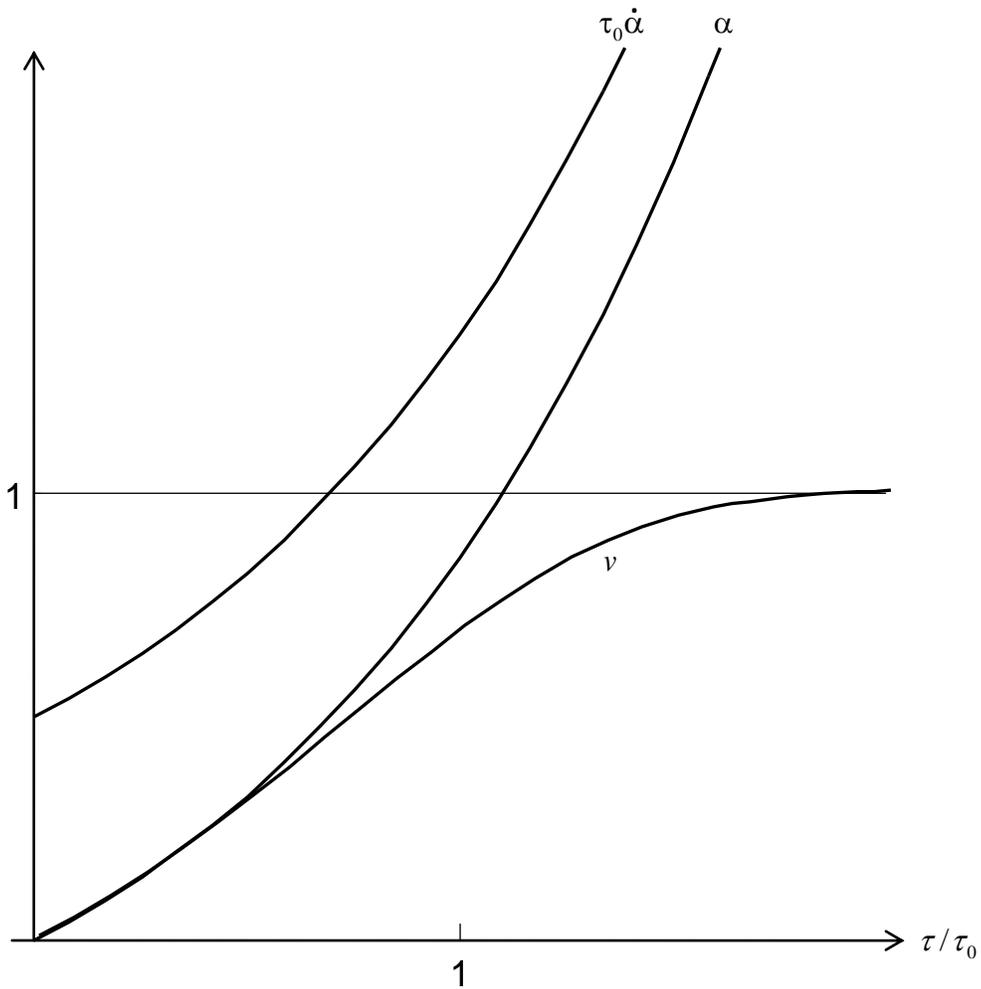

**Figure 2.** The proper acceleration, $\dot{\alpha}$, the velocity parameter, $\alpha$, and the velocity, $v = \tanh\alpha$, as functions of the proper time for a particle performing run away motion, starting from rest with positive acceleration. The quantity $\tau_0$ is the time taken by a light signal to travel a distance equal to two thirds of the particle's classical radius.

Noting that $\dot{\alpha}$ is the acceleration in the instantaneous inertial rest frame of the particle and that $k_L = m_0\tau_0$, we find the energies expressed by the rapidity utilizing, from Eq.(40), that $\dot{\alpha} = a_0 + \alpha/\tau_0$. The kinetic energy of the particle is

$$E_{kin} = m_0(\gamma - 1) = m_0(\cosh\alpha - 1) \,. \tag{41}$$

The radiation energy is

$$E_R = k_L \int_0^\tau \dot{\alpha}^2 \cosh\alpha \, d\tau = m_0(\alpha\sinh\alpha + a_0\tau_0\sinh\alpha - \cosh\alpha + 1) \,. \tag{42}$$

The Schott energy (also called acceleration energy) is

$$E_S = -k_L\gamma^4 va = -m_0\tau_0\dot{\alpha}\sinh\alpha = -m_0(\alpha + a_0\tau_0)\sinh\alpha \,. \tag{43}$$

The sum of the energies is constant and equal to the initial value zero.

Referring to Figure 1 we see that during the runaway motion the sum of the increase of kinetic energy and radiation energy comes from tapping a reservoir of Schott energy which is initially very close to the charge. This field energy becomes more and more negative, and the radius of the spherical surface limiting the region with Schott energy increases rapidly. Using Eqs.(21), (22) and (40) we find that it varies with the proper time of the particle as

$$T - T_{Q2} = \epsilon e^\alpha = \epsilon \exp\left[|a_0|\exp\left(\frac{\tau}{\tau_0}\right)\right] \,, \tag{44}$$

where $a_0$ is given in Eq.(38).

Next we consider the momenta. The momentum of the particle is

$$P_{kin} = m_0\gamma v = m_0\sinh\alpha \,. \tag{45}$$

The momentum of the radiation is

$$P_R = k_L \int_0^\tau \dot{\alpha}^2 \sinh\alpha \, d\tau = m_0(\alpha\cosh\alpha + a_0\tau_0\cosh\alpha - \sinh\alpha - a_0\tau_0) \,. \tag{46}$$

The Schott momentum (acceleration momentum) is

$$P_S = -k_L\gamma^4 a = -k_L\dot{\alpha}\cosh\alpha = -m_0(\alpha + a_0\tau_0)\cosh\alpha \,. \tag{47}$$

The sum of the momenta is constant and is equal to $-m_0 a_0 \tau_0$, which is the initial Schott momentum.

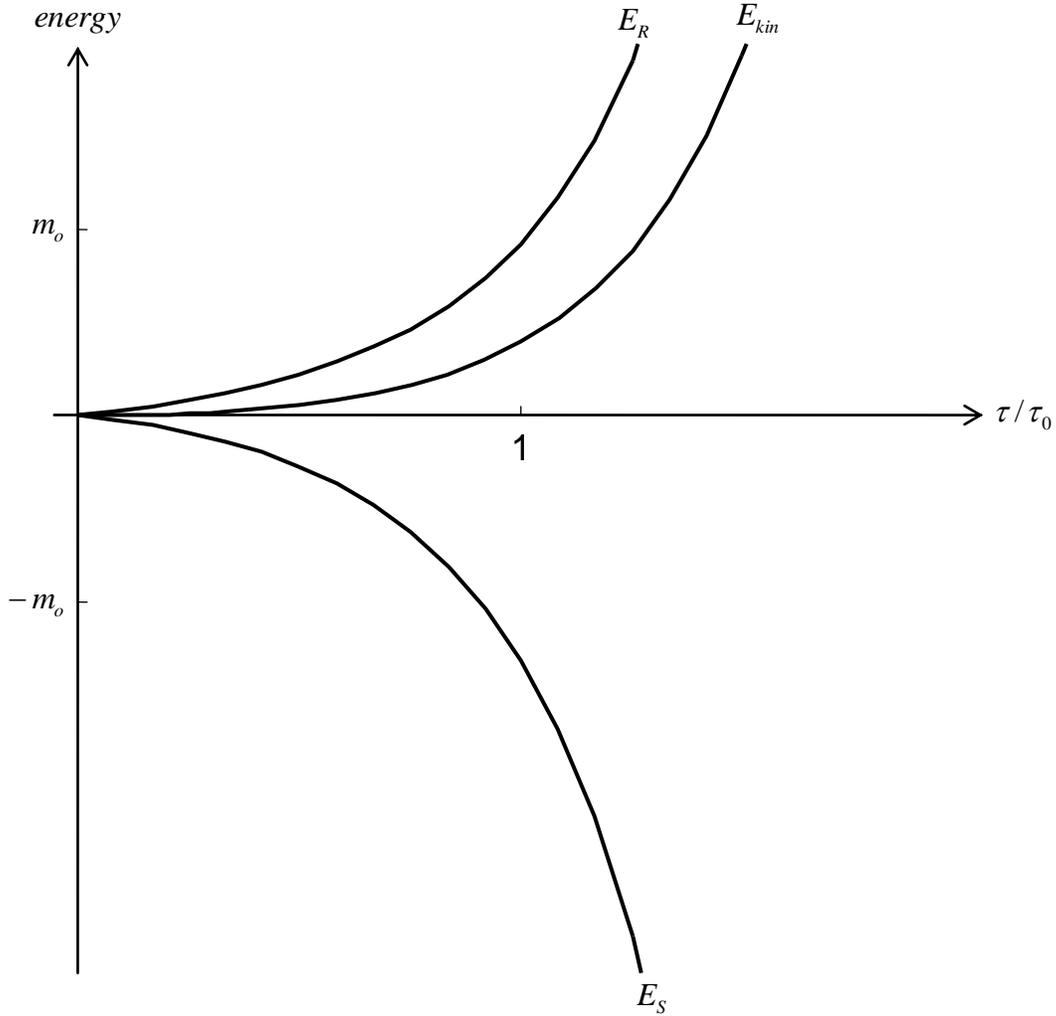

**Figure 3.** The energies of a particle and its electromagnetic field while the particle performs run away motion, as functions of $\tau/\tau_0$. Here $E_{kin}$ is kinetic energy, $E_R$ is radiated energy, and $E_S$ is Schott (or acceleration) energy.

The forces which are responsible for the increase in the momentum of the particle (internal forces) are the following (for rectilinear motion in general): The radiation reaction force,

$$\Gamma_R = -\frac{dP_R}{dT} = -k_L \dot{\alpha}^2 \tanh\alpha \ , \qquad (48)$$

and the acceleration reaction force,

$$\Gamma_A = -\frac{dP_S}{dT} = -\frac{1}{\cosh\alpha}\dot{P}_S = k_L(\ddot{\alpha} + \dot{\alpha}^2 \tanh\alpha) \ . \qquad (49)$$

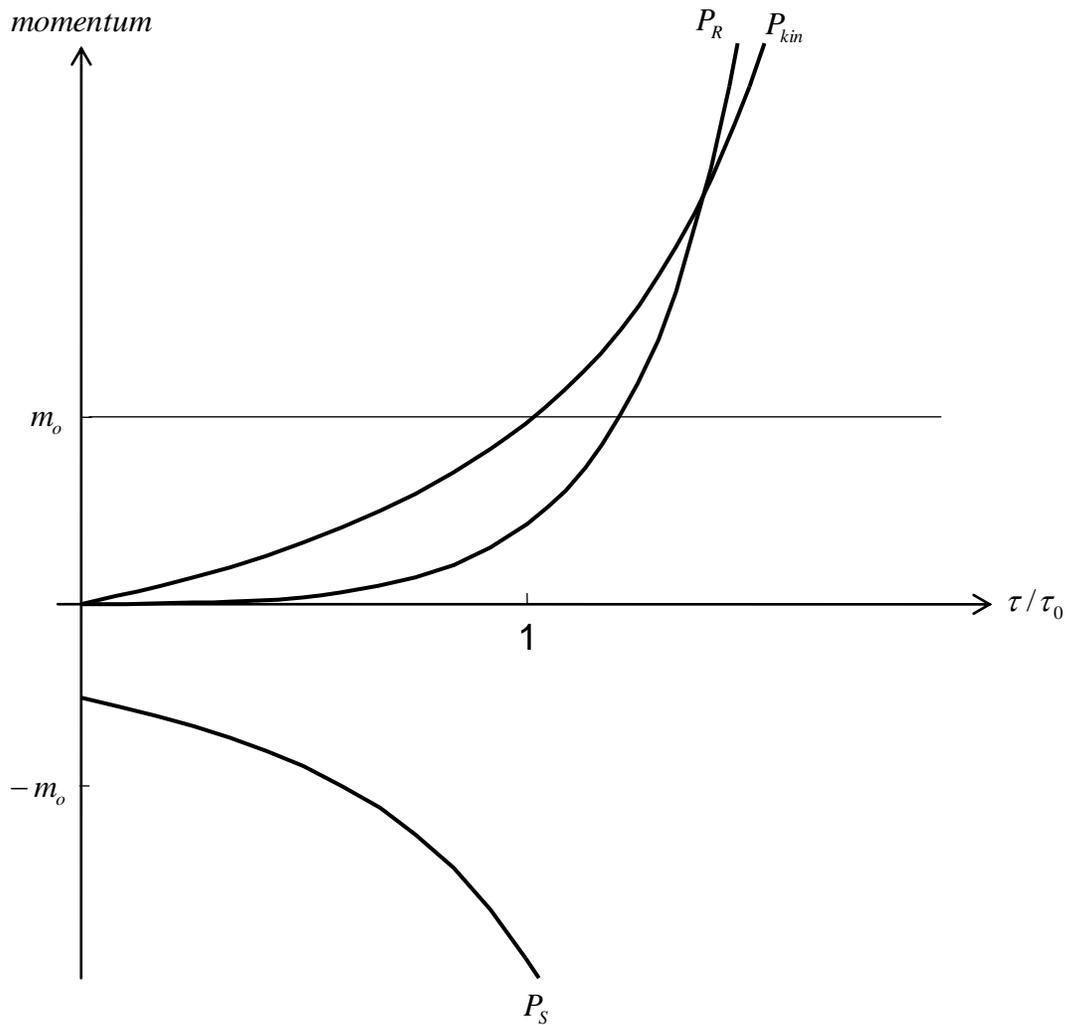

**Figure 4.** The momentum of a particle and its electromagnetic field while the particle performs run away motion, as functions of $\tau/\tau_0$. Here $P_{kin}$ is kinetic momentum, $P_R$ is radiated momentum, and $P_S$ is Schott (or acceleration) momentum.

The total field reaction force (also called the self force) is

$$\Gamma = \Gamma_R + \Gamma_A = k_L \ddot{\alpha} \ . \tag{50}$$

By means of Eq.(48) the forces are shown as functions of $\tau/\tau_0$ in Figure 5.

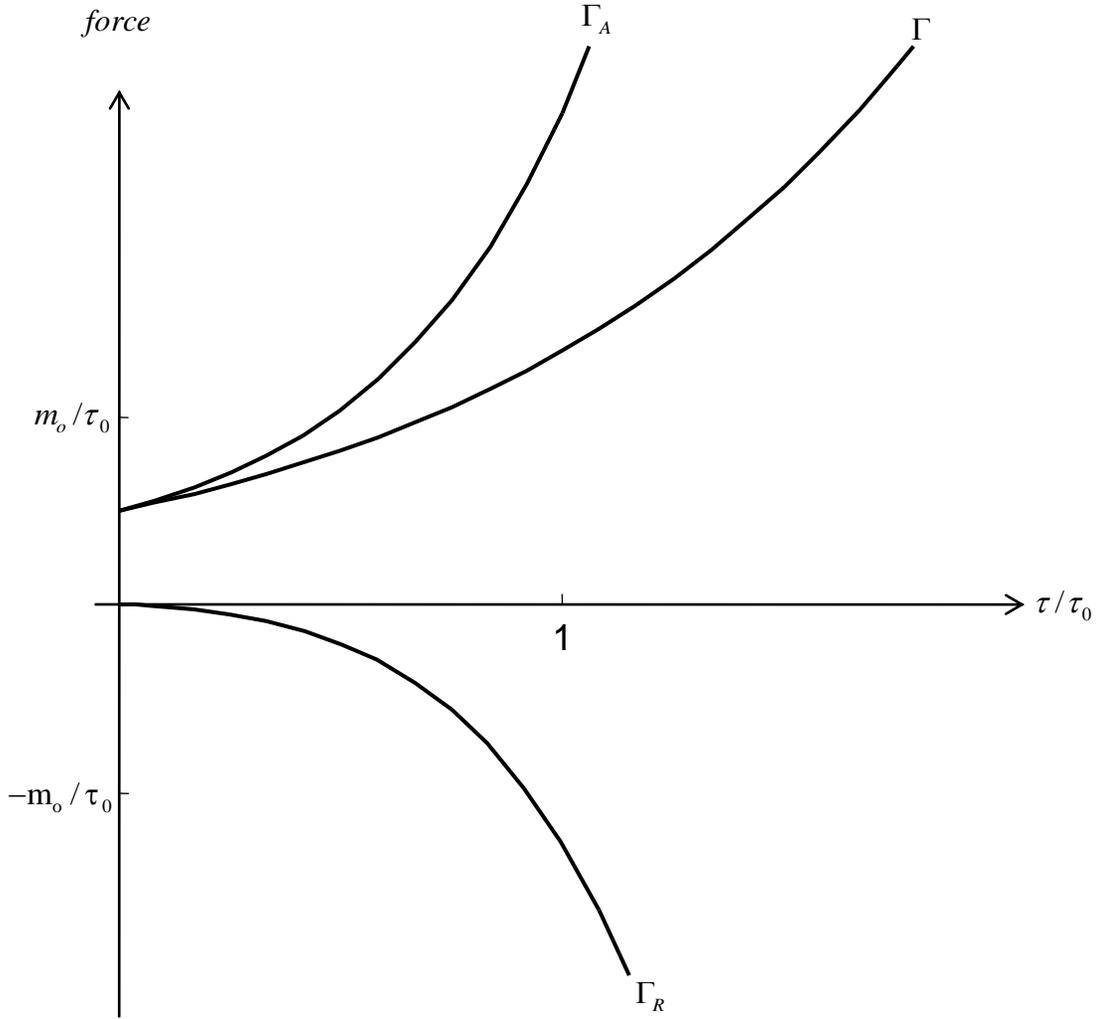

**Figure 5.** The forces due to the electromagnetic field of a particle acting on the particle while it performs run away motion, as functions of $\tau/\tau_0$. Here $\Gamma_R$ is the radiation reaction force, $\Gamma_A$ is the Schott (or acceleration) reaction force. Their sum is the field reaction force, $\Gamma = \Gamma_R + \Gamma_A$.

Eq.(48) shows that the radiation reaction force $\Gamma_R$ is a force that retards the motion, acting like friction in a fluid. The "push" in the direction of the motion is provided by the acceleration reaction force, which is opposite to the change of Schott momentum per unit time. This force is opposite to the direction of the external force, i.e. it has the same direction as the runaway motion.

There is a rather strange point here. We have earlier identified the Schott energy as a field energy localized close to the charge.[8] Yet, in the present case the Schott momentum is oppositely directed to the motion of the charge. This is due to the fact that the Schott energy is negative. Hence even if the Schott momentum has a direction opposite to that of the velocity of the charge, it represents a motion of negative energy in the same direction as that of the charge.

In general the Schott energy is

$$E_S = -k_L A^0 ,\qquad(51)$$

and the Schott momentum is

$$\mathbf{P}_S = -k_L \mathbf{A} ,\qquad(52)$$

Where $(A^0, \mathbf{A})$ is the 4-acceleration of the particle. From the relation $A^0 = \mathbf{v} \cdot \mathbf{P}_S$ we get $E_S = \mathbf{v} \cdot \mathbf{P}_S$. It follows that for rectilinear motion $\mathbf{v}$ and $\mathbf{P}_S$ are oppositely directed when $E_S$ is negative.

The Schott energy saves energy conservation for runaway motion of a radiating charge. Nevertheless, physicists doubt that runaway motion really exists. There is, however, no doubt that it is a solution of the LAD equation of motion of a charged particle. In this sense it is allowed, but maybe not everything that is allowed is obligatory. The physics equations seem to contain many possibilities that are not realized in our universe. And we seem to lack a criterion to eliminate those possibilities that do not exist physically. Hence, as a referee of the present article wrote: "One can only wonder why no runaways have ever been observed or why they could not be used as compact particle accelerators!"

### 5. Schott energy and radiated energy of a freely falling charge

The Rindler coordinates, $(t, x, y, z)$, of a uniformly accelerated reference frame are given by the following transformation from the coordinates $(T, X, Y, Z)$ of an inertial frame,

$$gt = \mathrm{artanh}(T/X) ,\quad x = \sqrt{X^2 - T^2}\qquad(53)$$

with inverse transformation

$$T = x\sinh(gt) ,\quad X = x\cosh(gt) .\qquad(54)$$

Here $g$ is a constant which shall be interpreted physically below.

Using Rindler coordinates, the line-element takes the form[17]

$$ds^2 = -g^2 x^2 dt^2 + dx^2 + dy^2 + dz^2 .\qquad(55)$$

In the Rindler frame the non vanishing Christoffel symbols are

$$\Gamma^x_{tt} = g^2 x \ , \quad \Gamma^t_{tx} = \Gamma^t_{xt} = 1/x \ . \tag{56}$$

The Rindler coordinates are mathematically convenient, but not quite easy to interpret physically. An observer at rest in a uniformly accelerated reference frame in flat spacetime experiences a field of gravity. From the geodesic equation,

$$\frac{du^\mu}{d\tau} + \Gamma^\mu_{\alpha\beta} u^\alpha u^\beta = 0 \ , \tag{57}$$

which is also the equation of motion of a freely moving particle, follows that acceleration of a free particle instantaneously at rest is

$$\frac{d^2 x}{dt^2} = -\Gamma^x_{tt} = -g^2 x \ . \tag{58}$$

Consider a fixed reference point $x = constant$ in the Rindler frame. It has velocity and acceleration

$$V = \frac{dX}{dT} = \tanh(gt) \ , \quad A = \frac{dV}{dT} = \frac{1}{x}\frac{1}{\cosh^3(gt)} \tag{59}$$

respectively, in the inertial frame. Hence, the acceleration of a reference point $x = constant$ at the point of time $t = 0$ is

$$A(0) = \frac{1}{x} \ . \tag{60}$$

This shows that the coordinate $x$ of the Rindler frame has dimension one divided by acceleration, and that it is equal to the inverse of the acceleration of the reference point that it represents at the point of time $t = 0$. The physical interpretation of the constant $g$ then follows from Eq.(57). It represents the acceleration of gravity experienced in the Rindler frame at the reference point having acceleration equal to $g$ relative to the inertial frame at the point of time $t = 0$.

The 4-velocity and the 4-acceleration of a particle moving along the $x$ −axis are

$$u^\mu = \frac{dx^\mu}{d\tau} = \gamma_R(1, v, 0, 0) \ , \quad \gamma_R = (g^2 x^2 - v^2)^{-1/2} \ , \tag{61}$$

$$a^\mu = \frac{du^\mu}{d\tau} + \Gamma^\mu_{\alpha\beta} u^\alpha u^\mu = \gamma_R^4 \left( a + g^2 x - \frac{2v^2}{x} \right)(v, g^2 x, 0, 0) \ , \tag{62}$$

where $v = dx/dt$ and $a = dv/dt$.

As seen from the expression (8) for the Abraham 4-force the field reaction force vanishes for a freely moving charge. Hence, such a charge falls with the same acceleration as a neutral particle. It has vanishing 4-acceleration and follows a geodesic curve. This is valid in a uniformly accelerated reference frame in flat spacetime, but not in curved spacetime.[18]

From the expression (62) is seen that in the present case the equation of motion may be written

$$x\frac{d^2x}{dt^2} - 2\left(\frac{dx}{dt}\right)^2 + g^2x^2 = 0 \ . \tag{63}$$

The solution of this equation for a particle falling from $x = x_0$ at $t = 0$ is

$$x = \frac{x_0}{\cosh(gt)} \ . \tag{64}$$

Hence

$$v = -gx_0\frac{\sinh(gt)}{\cosh^2(gt)} \ , \quad \gamma_R = \frac{\cosh^2(gt)}{gx_0} \ . \tag{65}$$

Saying that a charge radiates is not a reference independent statement. This conclusion has been arrived at in different ways.[19-23] M. Kretzschmar and W. Fugmann[20, 21] have generalized Larmor's formula to a form which is valid not only in inertial reference frames, but also with respect to accelerated frames. A consequence of their formula is that a charge will be observed to emit radiation only if it accelerates relative to the observer. Whether it moves along a geodesic curve is not decisive. A freely falling charge, i.e. a charge at rest in an inertial frame may be observed to radiate, and a charge acted upon by non-geodesic forces may be observed not to radiate.

In order to give a correct description of the radiation emitted by a charge valid in an accelerated frame of reference, one has to generalize the usual form of the Larmor formula for the radiated effect $\hat{P}_L$ valid in the orthonormal basis of an inertial frame,

$$\hat{P}_L = k_L\hat{a}^2 \ , \tag{66}$$

where $\hat{a}$ is the acceleration of the charge in an inertial frame. This formula says that an accelerated charge radiates, which is a misleading statement. It sounds as if whether a charge radiates or not, is something invariant that all observers can agree upon. However, that is not the case. An accelerated observer permanently at rest relative to an accelerated charge would not say that it radiates. The covariant generalization of the formula is

$$P_L = k_L A_\mu A^\mu \ . \tag{67}$$

Freely falling charges have vanishing 4-acceleration. Hence, this version of Larmor's formula seems to say that charges that are acted upon by non-gravitational forces radiate. As mentioned above this is not generally the case. In fact the formula above is not generally covariant. It is only Lorentz covariant because the components of the 4-acceleration are presupposed to be given with reference to an inertial frame in this formula.

Hirayama[22] recently generalized this formula to one which is also valid in a uniformly accelerated reference frame. He then introduced a new 4-vector which may be called the *Rindler 4-acceleration* of the charge. It is a 4-vector representing the acceleration of the charge *relative to the Rindler frame*, and has components

$$\alpha^\mu = a^\mu - \frac{1}{gx^2(gx+v)}(v, g^2x^2, 0, 0) \ . \tag{68}$$

where $a^\mu$ are the components in the Rindler frame of the 4-acceleration of the charge. For a freely falling charge $a^\mu = 0$. The generalised Larmor formula valid in a uniformly accelerated reference frame has the form

$$P = k_L g^2 x^2 \alpha_\mu \alpha^\mu \ , \tag{69}$$

and has been thoroughly discussed by Eriksen and Grøn.[17]

We shall now apply this formula to the charge falling freely from $x = x_0$ where it was instantaneously at rest. Then we need to calculate

$$\alpha_\mu \alpha^\mu = -g^2 x^2 (\alpha^t)^2 + (\alpha^x)^2 \ . \tag{70}$$

Inserting the expressions (64) and (65) for $x$ and $v$ in eq.(68) we obtain

$$\alpha^t = \frac{1}{gx_0^2} e^{gt} \sinh(gt) \cosh^2(gt) \ , \quad \alpha^x = -\frac{1}{x_0} e^{gt} \cosh^2(gt) \ , \tag{71}$$

which gives

$$\alpha_\mu \alpha^\mu = \frac{1}{x_0^2} e^{2gt} \cosh^2(gt) \ . \tag{72}$$

Inserting this into Eq.(69) we find the power radiated by the freely falling charge,

$$P = k_L g^2 e^{2gt} \ . \tag{73}$$

The radiated energy is

$$E_R = \int_0^t P dt = (k/2) g (e^{2gt} - 1) \ . \tag{74}$$

One may wonder where this energy comes from. A proton and a neutron will perform identical motions during the fall, although the proton radiates energy and the neutron not. As noted after Eq.(16), the answer is: The radiated energy comes from the Schott energy. The Schott energy is given by[14]

$$E_S = -k_L v \alpha^x \ . \tag{75}$$

Inserting the expressions (65) and (71) for $v$ and $\alpha^x$ respectively, we obtain for the Schott energy as a function of time,

$$E_S = -(k_L/2)g(e^{2gt} - 1) \ . \tag{76}$$

This shows that the radiation energy does indeed come from the Schott energy. Again the Schott part of the field energy inside the light front S in Figure 1 becomes more and more negative during the motion, and the region filled with Schott energy which is inside the light front S and outside the particle, has initially a vanishing volume, but increases rapidly in size.

### 6. Conclusion

Seemingly there is a problem with energy conservation connected with the LAD equation of motion of a radiating charge in combination with the Larmor formula for the effect of the radiation emitted by an accelerated charged particle, although a general analysis implies energy conservation for a dynamics based upon these equations.[23] The equation of motion has runaway solutions in which a charge accelerates and emits radiation even when it is not acted upon by any exterior force. Where does the increase of kinetic energy and radiation energy come from?

In the present article it has shown how the Schott energy provides both an increase of the kinetic energy of the particle and the energy it radiates. The Schott energy is the part of the electromagnetic field energy which is proportional to the acceleration of the charge, and for non-relativistic motion of the charge it is localized close to the charge.[5] But in the case of runaway motion the velocity of the charged particle approaches that of light and it is no longer true that the Schott energy is localized close to the particle. The Schott energy has the curious property that it can become increasingly negative, which makes it possible to use it as a sort of inexhaustible source of energy in the case of runaway motion.

Also the case of a freely falling charge in the gravitational field which exists in a uniformly accelerated reference frame in flat spacetime, is quite strange. The co-moving frame of the charge is an inertial frame in which it is permanently at rest. Obviously it does not radiate in this frame. Nevertheless it radiates as observed in the accelerated frame.[19] Again one may wonder: Where does the radiated energy come from? And again the answer is: It comes from the Schott energy.

We have here demonstrated how this comes about by calculating the radiated energy and the Schott energy as functions of time for runaway motion and for freely falling motion in a gravitational field. This provides an interesting application of the LAD equation that may be useful in the teaching of the electrodynamics of radiating charges. It has been shown that it is necessary to take the Schott energy into account in order to avoid apparent energy paradoxes in the theory of radiating charges based on the LAD equation.

The necessity of taking the Schott energy into account to save energy-momentum conservation may point to a problem with the LAD equation or the point particle model of a charge. Whereas there is a physical basis for the Schott energy in the electromagnetic field of a point charge, an energy that becomes negative without bound and supplies limitless radiation energy and kinetic energy of runaway solutions, may be a sign of the breakdown of the LAD equation.

**Acknowledgement**

I would like to thank the referees for useful comments that contributed to significant improvements of the article.